\documentclass[10pt,conference]{IEEEtran}
\usepackage[verbose]{cite}
\usepackage[verbose]{wrapfig}
\usepackage[dvipsnames]{xcolor}
\usepackage{cellspace}
\usepackage{moresize}
\usepackage{epsfig,psfrag,subfigure}
\usepackage[font=scriptsize,farskip=5mm]{subfig}
\usepackage{pstricks-add}
\usepackage{amsmath,amssymb}
\usepackage{multirow}
\usepackage{balance}

\begin{document}

\title{Consensus Problem under Diffusion-based Molecular Communication}
\author{ Arash Einolghozati, Mohsen Sardari, Ahmad Beirami, Faramarz Fekri\\
School of Electrical and
Computer Engineering, Georgia Institute of Technology, Atlanta, GA 30332\\
\texttt{Email:}~\{einolghozati, mohsen.sardari, beirami, fekri\}@ece.gatech.edu}

\maketitle


\begin{abstract}
We investigate the consensus problem in a network where nodes communicate via diffusion-based molecular communication (DbMC). In DbMC, messages are conveyed via the variation in the concentration of molecules in the medium. Every node acquires sensory information about the environment. Communication enables the nodes to reach the best estimate for that measurement, e.g., the average of the initial estimates by all nodes. We consider an iterative method for communication among nodes that enables information spreading and averaging in the network.  We show that the consensus can be attained after a finite number of iterations and variance of estimates of nodes can be made arbitrarily small via communication. 
\end{abstract}


\IEEEpeerreviewmaketitle

\section{Introduction}
\label{sec:intro}
There has been numerous evidence of the existence of a form of communication using molecules in nature. At the microorganism scale, molecular signals are used for communication and control among cells in living tissues.  Communication enables single cells to process sensory information about their environment (in a way similar to neural networks) and evaluate and react to chemical stimuli. The use of chemical signaling by living cells has been under extensive study. For example, it is known that some bacteria use a process named ``Quorum Sensing'' to estimate the density of their kind in the environment~\cite{Bassler1999,Bassler2002,Hammer2003,Henke2004,Mehta2009,Ng2009}. Quorum Sensing is a decentralized coordination process which allows bacteria to estimate the density of their population and regulate their behavior according to the estimated density. To estimate the local population density, bacteria release specific signaling molecules. These molecules are subject to diffusion process that would make the molecules drift away instead of accumulating in the bacteria vicinity. Therefore, the concentration near the bacteria tends to the average concentration in the medium. As the local density of bacteria increases, so will the density of molecules in the medium. Bacteria have molecules receptors that can estimate the molecules density and thus the bacteria population density. Bacteria use quorum sensing to coordinate actions (mostly energy expensive) that cannot be carried out by a single bacterium.
This phenomenon, captures most of the important components of a communication system in micro-scale. Arguably, the most dominating form of communication at the scale of microorganisms is Diffusion based Molecular Communication (DbMC), i.e., embedding the information in the alteration of the concentration of the molecules and rely on diffusion to transfer the information. 

Despite its importance, study of DbMC is still in infancy. Very recently, new theoretical frameworks are being developed for molecular communication~\cite{eckford, eckford2010,Pierobon2010a}. Broadly speaking, the purpose of understanding DbMC is three-fold. First, in the field of system biology. For example, the new field of system biology enables us to engineer microorganisms for certain objectives~\cite{Danino2010,Choudhary2010,Dressle2010}. Our study can help to model and manipulate biological networks comprised of such engineered microorganisms. Secondly, recently, there is a new trend of designing micro-scale networks of nano-scale devices to perform tasks similar to their biological counterparts~\cite{Akyildiz2010}. 
There is a large number of applications that such networks could apply to.  One may envision molecular based networks built using these nano-scale devices that can be deployed over or inside the human body to monitor glucose, sodium, and cholesterol levels, to detect the presence of different infectious agents, or to identify specific types of cancer. Such networks will also enable new smart drug administrative systems to release specific drugs inside the body with great accuracy and in a timely manner.
These networks are to operate in the environments similar to those of bacteria and other living organisms. Hence, the same principles hold for these networks and DbMC is the most promising form of communication in these bio-inspired networks. There is no need for complex computation in this type of communication which has been performed by bacteria for millions of years. Therefore, it can be done by primitive nano-scale devices that have limited computation capabilities. Third, understanding DbMC can help to find out whether there is any optimality in natural complex bio-systems.

In this work, we study the consensus problem in a network governed by DbMC. In a general consensus problem, nodes in a network communicate with each other to obtain the best estimate given their initial estimates. Average value of these initial estimates is considered to be an important measure that can be considered as a goal in a network. In this work, we capture the situation where a network of nodes (agents) must achieve a consistent opinion through information exchanges via molecular signaling with their neighbors.

The key element of this type of the consensus problem is diffusion. Diffusion describes the spread of particles through random motion from regions of higher concentration to regions of lower concentration. Every node has the capability of sensing the concentration of molecules in the environment and producing new molecules with desired rates. Molecules diffuse from the transmitter to the entire media. It is important to note that this type of molecular communication is different from other communication schemes where Brownian motion of single molecules is studied or molecules are directed toward specific directions. 

The rest of the paper is organized as follows: in Sec.~\ref{sec:back}, we capture the effect of diffusion by studying the Fick's second law of diffusion and show as to how the production of molecules by each node affects the concentration level sensed by other nodes in the network. In Sec.~\ref{sec:problem}, the consensus problem is presented and formulated in a matrix form. Then, in Sec.~\ref{sec:algorithm}, consensus problem is studied for compact and uniform networks and an iterative algorithm is proposed for the uniform type. In Sec.~\ref{sec:analysis}, the convergence of the proposed algorithm is verified and the rate of the convergence is discussed. Finally, in Sec.~\ref{sec:example}, the results in the previous sections is simulated and verified for a specific network.



\section{Background}
\label{sec:back}
\begin{figure*}
\begin{minipage}[b]{0.48\linewidth}
\includegraphics[height=0.8\linewidth, angle=-90]{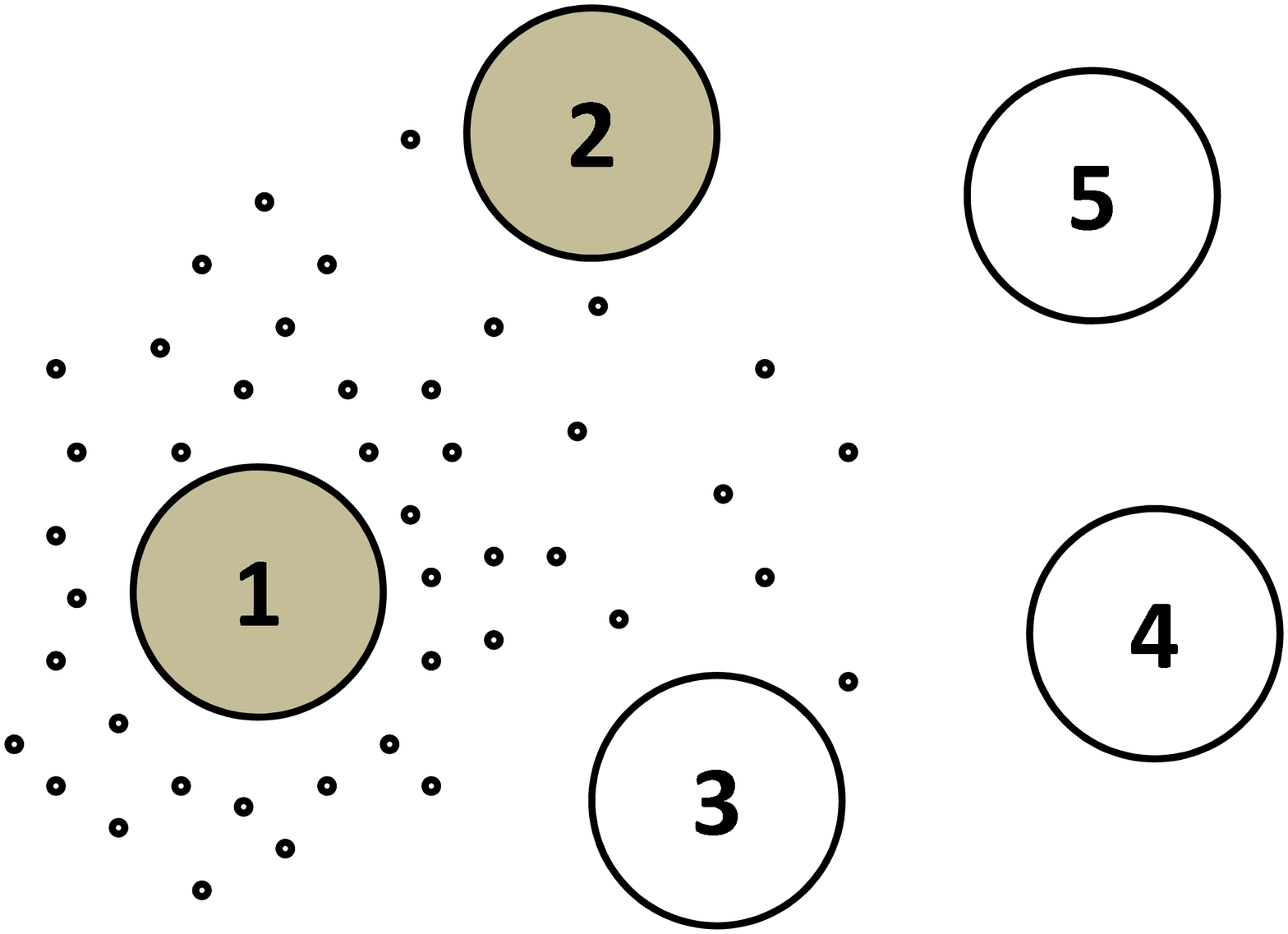}
\vspace{0.35in}
\caption{Node 1 Diffusing Molecules into a Network of 5 Nodes.}
\label{fig:network}
\end{minipage}
\hspace{0.03\linewidth}
\begin{minipage}[b]{0.48\linewidth}
\includegraphics[height =0.8\linewidth, angle = -90]{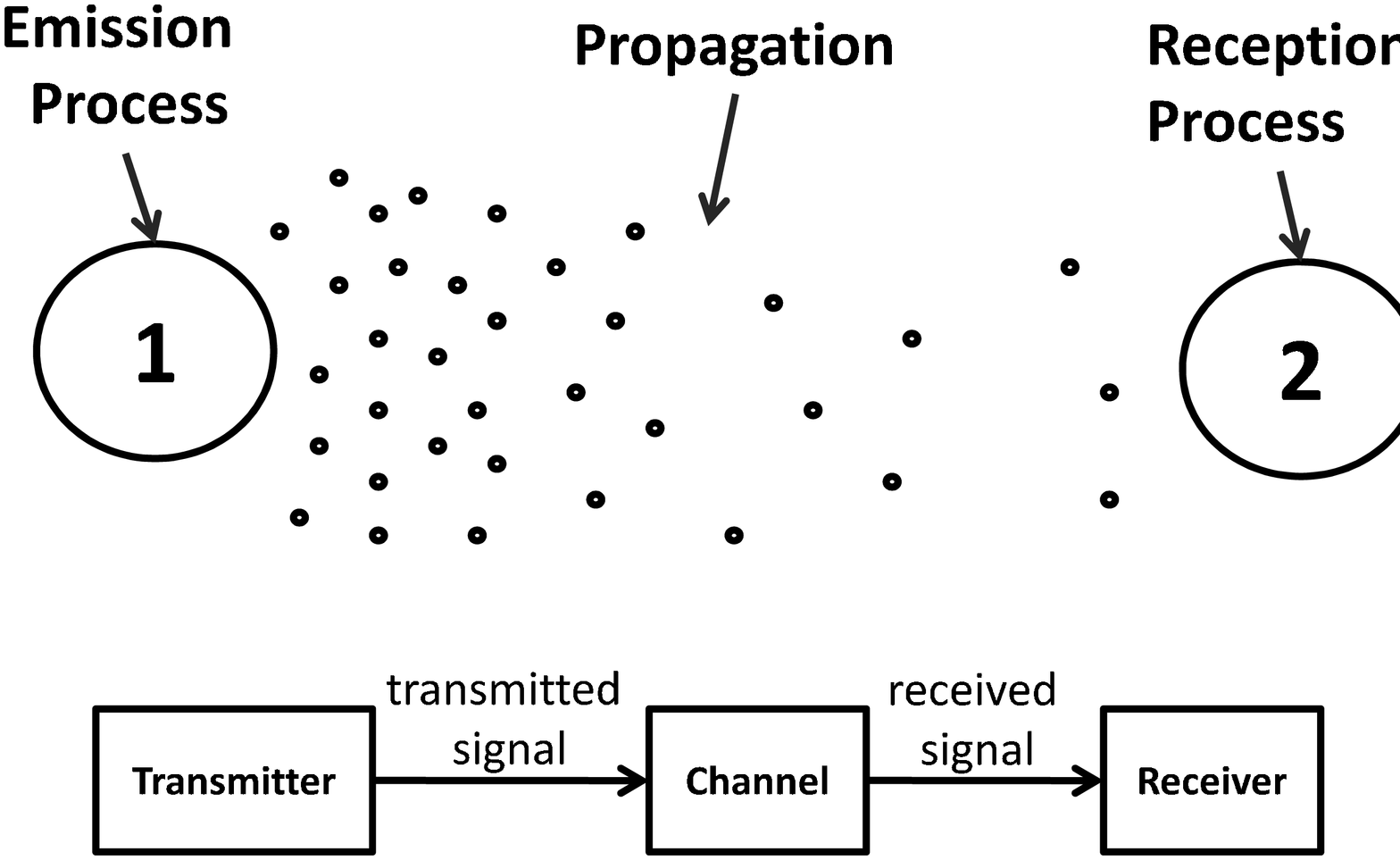}
\vspace{0.35in}
\caption{Model of Molecular Communication from node 1 to node 2.}
\label{fig:model}
\end{minipage}
\end{figure*}

The scheme for a typical molecular communication network is depicted in Fig.~\ref{fig:network}. The communication between the shaded nodes in the network is modeled in Fig.~\ref{fig:model}. Each agent has the capability of sensing the concentration of molecules in the medium and release molecules at a specific rate back into the medium. In other words, each agent in such networks is a transceiver. Channel is the medium that molecules are injected into and carried depending on the diffusion coefficient of the medium. In molecular communication, information is encoded into the variations of molecule concentration, e.g., it can be in the form of different concentration levels corresponding to Amplitude Modulation.


Two main features that distinguish the consensus problem under DbMC from that of the traditional electromagnetic communication are the broadcast nature of DbMC and lingering of the molecules in the shared medium. The molecules that are produced by nodes stay in the medium and change the concentration sensed by all the other nodes; there is no need for specialized routing which makes it fit for \emph{in-network} processing. Hence, the first step is to characterize the temporal and spatial variations of molecules in the channel which follows the general diffusion equations. According to Fick's second law of diffusion, the concentration of molecules $c(x,t)$ at position $x$ at time $t$ in an $m$-dimensional space (i.e. medium) is computed using the molecule production rate $r(x,t)$ at the source, as follows:

\begin{equation}
\label{eq:eq4}
 \frac {\partial c(x,t)}{\partial t} = D \nabla^2 c(x,t)+r(x,t).
\end{equation}
Here, $x$ is the distance of any point in the environment from the source (assuming the source is at the zero position) and $D$ is the diffusion coefficient of the medium. The impulse response of~(\ref{eq:eq4}), corresponding to $r(x,t)=\delta(x)\delta(t)$, is the Green's function $g_d(x,t)$ whose expression is as follows:
\begin{equation*}
 g_d(x,t)=\frac {1}{(4\pi Dt)^{\frac{m}{2}}} \exp{\left(-\frac{|x|^2}{4Dt}\right)}.
\end{equation*}
This impulse response is given for an $m$-dimensional medium. Note that diffusion in an $m$-dimensional space is equivalent to $m$ simultaneous 1-D diffusions.
Since the diffusion equation is a linear equation, the solution to~(\ref{eq:eq4}) for an arbitrary input $r(x,t)$, denoted by $c^*(x,t)$, can be calculated using  
\begin{equation*}
 c^*(x,t)=g_d(x,t)\otimes r(x,t),
\end{equation*}
where $\otimes $ denotes multi-dimensional convolution operation on $x$ and $t$.

In our setup, the only molecule producers are nodes. Therefore, we have $r(x,t)=F(t)\delta(x)$, where $F(t)$ is the molecule production rate. Hence, we will have
\begin{equation}
\label{eq:diffusion}
c^*(x,t) =\int^{\infty}_0 F(\tau)\frac {1}{(4\pi D(t-\tau))^{\frac{m}{2}}} \exp{\left( -\frac{x^2}{4D(t-\tau)}\right)} \,d{\tau}.
\end{equation}
This response is valid for open free medium where the only boundary conditions are at the transmitter. Note that in this model, we do not consider the delay due to the travel time of molecules between the nodes. That is we assume molecules reach the nodes instantly. This does not affect our analysis of consensus, because it only shifts the time that molecules are received at the nodes.

\section{Problem Statement}
\label{sec:problem}
We consider collaborative networks and refer to agents as network nodes. The goal is to spread the information about an event or any other variation through the network with minimum latency. The problem setup is such that each node in an $N$-node network initially has a measurement value.
These initial measurement values are assumed to be formed from estimating a specific parameter in the environment. These estimates are assumed to derive from a random variable. All the nodes try to obtain the best estimate for this random variable through the communication in the network. Node $i$ maps its estimate of the environment parameter into the level of molecule concentration $\rho_i$ corresponding to this estimate. Network nodes exchange their estimates via producing proper molecular concentrations in the medium, to arrive at consensus. We assume these estimated concentrations are drawn from the same distribution $N(\mu,\sigma_0^2)$, i.e., a Gaussian distribution with expected value of $\mu$ and variance of $\sigma_0^2$.

The best unbiased estimate for $\mu$ is the average of initial estimates of nodes, i.e. $\hat{\mu}=\rho_{av}=\frac{1}{N} \sum_{i=1}^N \rho_i$. This estimate has variance of $\frac{\sigma_0^2}{N}$ which can be arbitrarily small when $N$ is large enough.
Consensus is reached when estimate of each node approaches this value and the variance of their estimate approaches $\frac{\sigma_0^2}{N}$. We propose an algorithm that ensures all nodes to arrive at an estimate for the average consensus in finite amount of time. After evaluating the average of these concentrations, nodes can map their molecule production rate back into the average of the parameter that they measured in the environment.

In this network, nodes communicate with each other through diffusing molecules into the environment with different rates. These rates of molecule production change the concentration of molecules at vicinity of other nodes and convey the information from one node to another. Hence, each node needs to choose the best rate based on its estimate and convey this estimate to other nodes. Based on received molecule concentration and their previous estimate, nodes improve their estimates iteratively until they reach a consensus.

Assume node $i$ has an initial estimate $\rho_i$.  The goal is to obtain the average $\rho_{av} = \frac{1}{N} \sum_{i=1}^N \rho_i$ by every node. From~(\ref{eq:eq4}), the concentration of molecules $c_i(t)$ in the vicinity of node $i$ is given by
\begin{equation*}
 c_i(t)=\sum_{j=1}^N c^*_j(x_{ij},t) \quad \text{for}\quad i=1,2,\ldots,N.      
\end{equation*}
where $c^*_j(x_{ij},t)$ is the response at the position of node $i$ due to production by node $j$. Furthermore, $x_{ij}$ denotes the distance between nodes $i$ and $j$. In its classical communication analogue, we have a network of $N$ transceivers. Transmitter $i$ sends the signal $F_i(t)$ to $N$ receivers through $N$ different channels with the impulse response $g_d(x_{ij},t)$, $j\in \{1,2....,N\}$. At each receiver, the superposition of outputs from $N$ different channels is observed.

The $N$ equations above should be optimized with respect to $F_i(t)$ to produce the average of the initial estimates in the vicinity of each node at a specific time $T_0$. We assume node $i$ transmits molecules with a constant rate $F_i$ in the interval $[0,T_0]$. Therefore, at time $T_0$, the concentration of molecules at node $i$ is given by
\begin{equation}
\label{eq:superintegral}
 c_i(t) =\sum_{j=1}^N F_j X(i,j),   
\end{equation}
where 
\begin{equation}
\label{eq:elements}
X(i,j)= \int^ {T_0}_0 \frac {1}{(4\pi D(T_0-\tau))^{\frac{m}{2}}} \exp{\left(-\frac{|x_{ij}|^2}{4D(T_0-\tau)}\right)} \,d{\tau}. 
\end{equation}
Note that $X(i,j)$ only depends on the geometry of the network.

Let $F$ be a vector consisting of the production rate of all the nodes. Likewise, let ${\bf c}$ be the vector of concentration of molecules at vicinity of the network nodes. Then, we have
\begin{equation}
\label{eq:matrix}
{\bf c} = {\bf X}F.
\end{equation}
In~(\ref{eq:matrix}), ${\bf X}$ is a symmetric matrix whose entries $X(i,j)$ are calculated according to~(\ref{eq:elements}). If we set every entry of ${\bf c}$ as $\rho_{av}$, we will get a linear system of equations that can be solved to obtain $F_i$, the $i^{th}$ element of $F$. If ${\bf X}$ is invertible, there would be a unique solution for $F$. The important issue here is that each agent chooses $F_i$ based solely on its estimate whereas the estimates of other nodes are unknown for the agent. Hence, $F_i$ in~(\ref{eq:matrix}) can depend only on $\rho_i$ but the general solution for $F$ does not necessarily satisfy this condition. Therefore, we propose an iterative solution for reaching the consensus while imposing this dependency constraint.

\section{Proposed Algorithm}
\label{sec:algorithm}

We consider a discrete model of time epochs of length $T_0$ for analyzing the network. The length of an epoch depends on the topology of the network. In particular, it depends on the effective radius $R$ of each node (which is explained later) and the diffusion coefficient of the medium $D$. We can assume $T_0=k\frac{R^2}{D}$ where $k$ is a constant. We arbitrarily choose $k$ to be equal to one. Although our analysis is valid for other values of $k$. In each time epoch, node $i$ emits molecules with a constant rate $F_i$. At the beginning of each time epoch, each node optimizes $F_i$ based on its previous estimate and measurement of the molecular concentration in the environment, such that the estimate of all nodes becomes closer to $\rho_{av}$. Hence, at the end of each interval $(t=T_0)$, the concentration at node $i$ can be obtained by~(\ref{eq:matrix}).

Note that after each interval, nodes need to wait for a specific amount of time, which will be again in the form of $k\frac{R^2}{D}$, before releasing the molecules for the next epoch. This waiting interval is needed to allow the molecules in the environment diffuse away or equivalently to reset the channel. For the rest of the paper, an iteration includes both the injection of molecules and the waiting intervals.  

{\bf Case I}: First, we study the special case of compact networks. In such networks, nodes are in vicinity of each other and approximately sense the same concentration of molecules within the radius of the network, i.e. $c_i = c_j$, $\forall i,j$. Molecules produced by each node would have the same share in making up the concentration in the network radius. Therefore, the matrix equation in~(\ref{eq:matrix}) results in a single equation for the common concentration of nodes $c(T_0)$ at the center of the network at time $T_0$:

\begin{eqnarray}
c(T_0) &=& \sum_{j=1}^N \left[ \int^{T_0}_0   \frac {F_j}{(4\pi D(T_0-\tau))^{\frac{m}{2}}} \right. \nonumber\\
&& \qquad \left. \exp{\left(-\frac{|x_j|^2}{4D(T_0-\tau)}\right)} \,\mathrm{d} {\tau}\right] \nonumber \\
\label{eq:compact}
&=& \sum_{j=1}^N F_j X_j.
\end{eqnarray}
Here, $x_j$ is the distance between the node $j$ and the center of the network (note that we choose the center for the notation simplification). Furthermore, $X_j$ is the result of integration in~(\ref{eq:compact}) which depends only on $x_j$ and the network constants.

Equation~(\ref{eq:compact}) implies that by setting $F_j= \frac{\rho_j}{X_j N}$, we can equate the concentration $c(T_0)$ to $\rho_{av}=\frac{1}{N}\sum_{j=1}^N \rho_j$. We observe that the previously mentioned dependency constraint is taken into account. In this scenario, all the nodes are able to observe each other. Hence, by setting appropriate rates, they can reach the consensus in one step without any further iterations. In the following, we study the more challenging case that the network is extended and the concentration of molecules differs at the vicinity of each node. In addition, nodes may not be able to interact with all the other nodes. In this case, the information is conveyed from one side of the network to another by nodes acting as relays.

{\bf Case II}: We consider a network with nodes that are uniformly deployed in an $m$-dimensional medium. In a sufficiently large network, each node observes the same relative distances to the other nodes in the network (except for the nodes on the boundary of the network whose effects are negligible when the number of nodes is large enough). Since the network is extended, each node has effective communications with only a specific number of nodes. To obtain this number, we consider a circle of radius $r$ around each node which includes the nodes that it can have effective communications. The number of nodes at distance $r$ from node $i$ increases linearly with $r$. However, based on~(\ref{eq:diffusion}), the effect they have on node $i$ decreases as $e^{-r^2}$. Hence, the collective effect of all the nodes at distance $r$ varies as a factor of $r e^{-r^2}$ which approaches zero when $r$ becomes large. By setting a threshold $\epsilon$ for this factor, we can find $R$, the effective radius around each node. Therefore, the number of nodes $N^\prime$ that each node is able to interact with, is equal to $\pi R^2 d$ in a 2-$D$ medium, where $d$ is density of nodes in the medium. 

Based on the above discussion, the matrix ${\bf X}$, will be an $(N\times N)$ symmetric matrix where the columns are the permutation of each other. In addition, at each column (or equivalently at each row), there are $N^\prime$ nonzero elements whereas the rest of $N-N^\prime$ elements are zero. These zero elements correspond to the nodes that the node cannot interact directly. In this model, the effect of nodes on the boundary is disregarded. We note that in a general case, ${\bf X}$ would be a general symmetric matrix and the analysis will be more difficult. However, in the uniformly distributed case, we can propose an iterative method that ensures the reduction of variance in each iteration and convergence to the average value. 

Let $ \boldsymbol {\rho}(n)$ be the estimate vector of nodes at epoch $n$ and $\ \rho_i(n)$ be the $i^{th}$ element of the vector. In addition, let ${\bf c}(n)$ denotes the vector of molecule concentrations at the vicinity of nodes at epoch $n$. Assume each node has an initial estimate $\rho_i(0)$. After each epoch, nodes update their estimate based on the molecules received from other nodes. Suppose $S_i$ to be the sum of entries in the $i^{th}$ column of ${\bf X}$. Because of the isotropic setting, $S_i$ is the same for all columns, denoted by $S$. The nodes set their molecule production rate as $F(n)=\frac{\boldsymbol{\rho}(n)} {S}$ (we assume that each node is aware of the location of other nodes at its effective radius and hence can compute $S$). Hence, based on~(\ref{eq:matrix}), after communication among nodes, we will have
\begin{equation}
\label{eq:extended}
 {\bf c}(n)={\bf X} \frac{\boldsymbol{\rho}(n)}{S}=\frac{\bf X}{S} \boldsymbol{\rho}(n).
\end{equation}

We observe that sum of each row or column in the matrix $\tilde{\bf X}=\frac{\bf X}{S}$ is equal to one. Hence $\tilde{\bf X}$ is a doubly stochastic matrix. We assume that each node $i$ updates its estimate by putting $\rho_i(n+1)=c_i(n)$. Thus, we have an iterative equation for estimate of nodes.
\begin{equation}
\label{eq:iteration}
\boldsymbol{\rho}(n)=\tilde{\bf X}\boldsymbol{\rho}(n-1).
\end{equation}
In the following, we examine the convergence of~(\ref{eq:iteration}). 

\section{Convergence Analysis}
\label{sec:analysis}
First, we verify that the iterative algorithm proposed in~(\ref{eq:iteration}) results in an unbiased estimate for $\mu=\textsf{E}[\rho_i(0)]$ where $\mu$ is the expected value of the Gaussian distribution that the initial estimate is from. We denote by $\textsf{E}[\boldsymbol{\rho}(n)]$ the vector of expected values of the estimates at iteration $n$. We know that $\textsf{E}[\boldsymbol{\rho}(0)]=\mu{\bf 1}$, where ${\bf 1}$ is the $(N\times 1)$ vector whose elements are all one. Since the matrix $\tilde{\bf X}$ is assumed to be constant during the iterations, $\textsf{E}[\boldsymbol{\rho}(n)]$ is given by
\begin{equation}
\label{eq:average}
\textsf{E}[\boldsymbol{\rho}(n)]=\textsf{E}[\tilde{\bf X}\boldsymbol{\rho}(n-1)]=\tilde{\bf X}\textsf{E}[\boldsymbol{\rho}(n-1)].
\end{equation}
Since the sum of each row of $\tilde{\bf X}$ is equal to zero, $\tilde{\bf X} \textsf{E}[\boldsymbol{\rho}(0)]= \textsf{E}[\boldsymbol{\rho}(0)]$. Hence, by continuing the chain in~(\ref{eq:average}), we conclude that $\textsf{E}[\boldsymbol{\rho}(n)]=\textsf{E}[\boldsymbol{\rho}(0)]=\mu{\bf 1}$. This implies that the estimate of nodes at each step $n$ is an unbiased estimate for the initial parameter.

In order to study the variance of the estimates in each iteration, we need to elaborate on some characteristics of the matrix $\tilde{\bf X}$. This doubly stochastic matrix resembles the transition matrix in a Markov chain. We observe that this transition matrix is aperiodic because the entries on the main diagonal, which shows the effect of each node on its own, are nonzero. It is also irreducible because the graph representation of this matrix is connected and we can reach other nodes from each node. Hence, we use the Perron-Frobenius Theorem~\cite{prob}, regarding the eigenvalues ($\lambda$) of $\tilde{\bf X}$:
\begin{enumerate}
\item  $\lambda_1 = 1$ and it is a singular root,
\item $|\lambda_N| \leq |\lambda_{N-1}| \dots \leq |\lambda_2| < 1$.
\end{enumerate}
Since the sum of each row is one, the eigenvector corresponding to $\lambda_1$ is a uniform vector which is the consensus vector. The normalized form of this vector is $v_1= \frac{1}{\sqrt{N}}{\bf 1}$. The eigenvalue decomposition of $\tilde{\bf X}$ is given by
\begin{equation}
\label{eq:decompose}
\tilde{\bf X}=Q \Lambda {Q}^{-1}=Q \Lambda {Q}^{T}.
\end{equation}
Here, $Q$ is a matrix that whose columns are the eigenvectors of $\tilde{\bf X}$ and $\Lambda$ is a diagonal matrix containing the eigenvalues of $\tilde{\bf X}$. The result in~(\ref{eq:decompose}) comes from the fact that $\tilde{\bf X}$ is a symmetric matrix. Hence, eigenvectors are orthogonal to each other and we have ${\tilde{\bf X}}^{-1}= {\tilde{\bf X}}^T$. Based on~(\ref{eq:decompose}), we have $\tilde{\bf X}^k=Q \Lambda^k {Q}^T $. Therefore, the estimate of nodes at iteration $n$ is given by
\begin{eqnarray}
\label{eq:estimate}
\boldsymbol{\rho}(n)&=&\tilde{\bf X} \boldsymbol{\rho}(n-1) \nonumber\\
&=&\tilde{\bf X}^n \boldsymbol{\rho}(0) \nonumber\\
&=&Q \Lambda^n Q^T \boldsymbol{\rho}(0).
\end{eqnarray}
Since all the eigenvalues except for $\lambda_1=1$ are smaller than one, all the entries in the main diagonal of matrix $\Lambda$ approaches zero except for $\Lambda_{11}=1$. Hence, we have
\begin{eqnarray}
\label{eq:limit}
\lim_{n \to \infty} \boldsymbol{\rho}(n) &=& v_1 v_1^T \boldsymbol{\rho}(0) \nonumber\\
&=& \frac{1}{N} \left(\sum_{i=1}^N \rho_i(0)\right) {\bf 1} \nonumber\\
&=& \rho_{av} {\bf 1}
.\end{eqnarray}

As we see in~(\ref{eq:limit}), after sufficient number of iterations, the estimate of nodes approaches to the average of initial beliefs. In order to quantify the rate of convergence, we look into the variance of estimates of nodes. As mentioned before, the smallest variance that can be achieved is $\frac {\sigma_0^2}{N}$ which corresponds to the average value. The covariance matrix $Cov(n)$ of the estimates at iteration $n$ is given by
\begin{eqnarray}
\label{eq:variance}
Cov(n) &=& \textsf{E}\left[ (\boldsymbol{\rho}(n)-\textsf{E}[\boldsymbol{\rho}(n)]) (\boldsymbol{\rho}(n)-\textsf{E}[\boldsymbol{\rho}(n)])^T\right] \nonumber\\
&=& \tilde{\bf X}^n \textsf{E}\left[\boldsymbol{\rho}(0)-\textsf{E}[\boldsymbol{\rho}(0)]) \right. \nonumber\\
&&  \qquad \left. (\boldsymbol{\rho}(0)-\textsf{E}[(\boldsymbol{\rho}(0)])^T\right] (\tilde{\bf X}^T)^n \nonumber\\
&=&\tilde{\bf X}^n Cov(0) \tilde{\bf X}^n.
\end{eqnarray}
where the fact that all powers of a symmetric matrix are also symmetric is used. Since the initial estimates are considered to be independent of each other, $Cov(0)=\sigma_0^2 {\bf I}_{N\times N}$ where ${\bf I}$ is the identity matrix. Hence, from~(\ref{eq:variance}), $Cov(n)= \sigma_0^2 \tilde{\bf X}^{2n}$. We denote the diagonal elements of the matrix $Cov(n)$ by $Cov_{ii}(n)$ which gives the variance of the estimate of each node at iteration $n$. For large $n$, we consider only the effect of $\lambda_2$, the second largest eigenvalue of $\tilde{\bf X}$. Thus, by using the decomposition in~(\ref{eq:decompose}), we have
\begin{equation}
\label{converge}
Cov_{ii}(n)=\sigma_0^2 (\frac{1}{N}+ v_{2i}^2 \lambda_2^{2n}),
\end{equation}
where $v_{2i}$ denotes the $i^{th}$ element of eigenvector $v_2$ corresponding to $\lambda_2$. Since the norm of eigenvector $v_2$ is one, then $v_{2i}\leq 1, \,  i \in {1,2....,N}$. Hence, the variance of the estimate of node $i$ at iteration $n$ is obtained as 
\begin{equation}
\label{eq:bound}
\sigma_i^2(n) \leq \sigma_0^2 (\frac{1}{N}+ \lambda_2^{2n}) \quad i \in {1,2....,N}.
\end{equation}
By deploying more iterations, the upper-bound in~(\ref{eq:bound}) can become arbitrarily close to $\frac{\sigma_0^2}{N}$ which approaches to zero in a network with large number of nodes. It is obvious that smaller $\lambda_2$ will result in a faster convergence. It can be proved that when matrix $\tilde{\bf X}$ becomes more sparse, i.e. the columns and the rows contain more zeros, $\lambda_2$ becomes closer to one. This can be explained by the fact that a more sparse $\tilde{\bf X}$ means a more extended network. Hence, more number of relays are needed to spread the information in the network and hence, reaching consensus will be more time consuming. In particular when $\tilde{\bf X}$ becomes a diagonal matrix, $\lambda_2$ is equal to one and $\sigma_i^2$ does not converge to $\frac{1}{N}\sigma_0^2$. This case is equivalent to the scenario in which none of the nodes are able to communicate with each other and their initial beliefs cannot be improved.

\begin{figure}
\centering
\vspace{-0.05in}
\includegraphics[width = .9\linewidth]{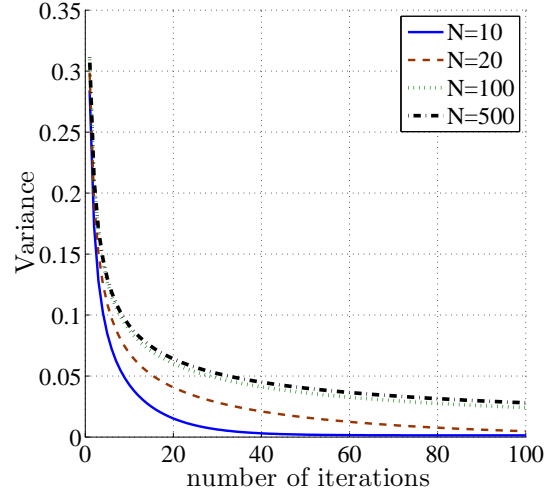}
\caption{Convergence of the iterative algorithm versus the number of nodes in the network.}
\vspace{-0.15in}
\label{fig:converge}
\end{figure}


\section{Case Study}
\label{sec:example}
In this section, we present an example to verify our results in the previous section. We assume a network of $N$ nodes that are placed on a line where the distance between two successive nodes is a constant $a$. This network would satisfy the uniform network condition in Sec.~\ref{sec:algorithm} as long as this line extends to infinity in both ends. For a realistic situation, we assume $N$ to be finite. We assume nodes communicate with each other through $2$-D DbMC in the medium and the size of nodes is small such that it does not interfere with diffusion of molecules. 

We compute the elements of matrix {\bf X} based on~(\ref{eq:elements}) and by normalizing the columns, we obtain $\tilde{\bf X}$. Since, the boundary nodes are taken into account, the symmetric matrix of $\tilde{\bf X}$ would not be doubly stochastic. We compute the variance of columns of powers of $\tilde{\bf X}$ and verify the convergence of the matrix that governs the convergence of the estimates. The results for variant $N$ and fixed effective radius, i.e. fixed $N'$, is shown in Fig.~\ref{fig:converge}.  We assumed $T_0$ to be $\frac{a^2}{D}$ and $N'$ equal to $5$. In this plot, sum of the variance of columns is depicted versus the number of iterations. As we observe in the graph, as $N$ becomes larger, $\tilde{\bf X}$ becomes more sparse and the rate of convergence decreases since the second largest eigenvalue becomes larger. 

\section{Conclusion}
In this paper, we studied the consensus problem in a network that employs DbMC for communication between nodes. We first considered the simple case of compact networks and then extended the concept to more general distributed networks. In particular, we considered a network with uniformly distributed nodes in the medium. Then, we developed a protocol as to how each node should behave to arrive at an unbiased estimate. The resulting protocol is an iterative scheme which the variance of the estimates can be reduced arbitrarily. Future works include the general case of networks and the case that nodes do not have complete information about the other nodes in their vicinity.

\balance
\bibliographystyle{IEEEtran}
\bibliography{CISS_2011_Consensus_v2}

\begin{thebibliography}{10}
\providecommand{\url}[1]{#1}
\csname url@samestyle\endcsname
\providecommand{\newblock}{\relax}
\providecommand{\bibinfo}[2]{#2}
\providecommand{\BIBentrySTDinterwordspacing}{\spaceskip=0pt\relax}
\providecommand{\BIBentryALTinterwordstretchfactor}{4}
\providecommand{\BIBentryALTinterwordspacing}{\spaceskip=\fontdimen2\font plus
\BIBentryALTinterwordstretchfactor\fontdimen3\font minus
  \fontdimen4\font\relax}
\providecommand{\BIBforeignlanguage}[2]{{%
\expandafter\ifx\csname l@#1\endcsname\relax
\typeout{** WARNING: IEEEtran.bst: No hyphenation pattern has been}%
\typeout{** loaded for the language `#1'. Using the pattern for}%
\typeout{** the default language instead.}%
\else
\language=\csname l@#1\endcsname
\fi
#2}}
\providecommand{\BIBdecl}{\relax}
\BIBdecl

\bibitem{Bassler1999}
B.~L. Bassler, ``\BIBforeignlanguage{eng}{How bacteria talk to each other:
  regulation of gene expression by quorum sensing.}''
  \emph{\BIBforeignlanguage{eng}{Curr Opin Microbiol}}, vol.~2, no.~6, pp.
  582--587, Dec 1999.

\bibitem{Bassler2002}
------, ``\BIBforeignlanguage{eng}{Small talk. cell-to-cell communication in
  bacteria.}'' \emph{\BIBforeignlanguage{eng}{Cell}}, vol. 109, no.~4, pp.
  421--424, May 2002.

\bibitem{Hammer2003}
B.~K. Hammer and B.~L. Bassler, ``\BIBforeignlanguage{eng}{Quorum sensing
  controls biofilm formation in vibrio cholerae.}''
  \emph{\BIBforeignlanguage{eng}{Mol Microbiol}}, vol.~50, no.~1, pp. 101--104,
  Oct 2003.

\bibitem{Henke2004}
\BIBentryALTinterwordspacing
J.~M. Henke and B.~L. Bassler, ``\BIBforeignlanguage{eng}{Bacterial social
  engagements.}'' \emph{\BIBforeignlanguage{eng}{Trends Cell Biol}}, vol.~14,
  no.~11, pp. 648--656, Nov 2004. [Online]. Available:
  \url{http://dx.doi.org/10.1016/j.tcb.2004.09.012}
\BIBentrySTDinterwordspacing

\bibitem{Mehta2009}
\BIBentryALTinterwordspacing
P.~Mehta, S.~Goyal, T.~Long, B.~L. Bassler, and N.~S. Wingreen,
  ``\BIBforeignlanguage{eng}{Information processing and signal integration in
  bacterial quorum sensing.}'' \emph{\BIBforeignlanguage{eng}{Mol Syst Biol}},
  vol.~5, p. 325, 2009. [Online]. Available:
  \url{http://dx.doi.org/10.1038/msb.2009.79}
\BIBentrySTDinterwordspacing

\bibitem{Ng2009}
\BIBentryALTinterwordspacing
W.-L. Ng and B.~L. Bassler, ``\BIBforeignlanguage{eng}{Bacterial quorum-sensing
  network architectures.}'' \emph{\BIBforeignlanguage{eng}{Annu Rev Genet}},
  vol.~43, pp. 197--222, 2009. [Online]. Available:
  \url{http://dx.doi.org/10.1146/annurev-genet-102108-134304}
\BIBentrySTDinterwordspacing

\bibitem{eckford}
K.~V. Srinivas, R.~S. Adve, and A.~W. Eckford, ``Molecular communication in
  fluid media: The additive inverse gaussian noise channel,''
  arXiv:1012.0081v2.

\bibitem{eckford2010}
A.~W.~E. Sachin~Kadloor, Raviraj S.~Adve, ``Molecular communication using
  brownian motion with drift,'' arXiv:1006.3959v1.

\bibitem{Pierobon2010a}
M.~Pierobon and I.~F. Akyildiz, ``A physical end-to-end model for molecular
  communication in nanonetworks,'' \emph{IEEE Journal of Selected Areas in
  Communications}, vol.~28, pp. 602--611, 2010.

\bibitem{Danino2010}
T.~Danino, O.~Mondragón-Palomino, L.~Tsimring, and J.~Hasty, ``A synchronized
  quorum of genetic clocks.'' \emph{Nature}, vol. 463, no. 7279, pp. 326--330,
  Jan 2010.

\bibitem{Choudhary2010}
\BIBentryALTinterwordspacing
S.~Choudhary and C.~Schmidt-Dannert, ``\BIBforeignlanguage{eng}{Applications of
  quorum sensing in biotechnology.}'' \emph{\BIBforeignlanguage{eng}{Appl
  Microbiol Biotechnol}}, vol.~86, no.~5, pp. 1267--1279, May 2010. [Online].
  Available: \url{http://dx.doi.org/10.1007/s00253-010-2521-7}
\BIBentrySTDinterwordspacing

\bibitem{Dressle2010}
F.~Dressle and O.~B. Akan, ``A survey on bio-inspired networking,''
  \emph{Computer Networks}, vol.~56, pp. 881--900, 2010.

\bibitem{Akyildiz2010}
I.~Akyildiz and J.~M. Jornet, ``Nano communication networks,''
  \emph{NanoComNet}, vol.~1, pp. 3--19, 2010.

\bibitem{prob}
G.~R. Grimmett and D.~R. Stirzaker, \emph{{Probability and Random
  Processes}}.\hskip 1em plus 0.5em minus 0.4em\relax Oxford University Press,
  2001.

\end{thebibliography}

\end{document}